\documentclass[review,12pt]{elsarticle}

\usepackage{amssymb}

\usepackage{color}

\begin{document}

\begin{frontmatter}



\title{Vulnerability Analysis of PAP for RFID Tags}


\author{Mu'awya Naser}
\address{School of Computer Sciences, Main Campus, Universiti Sains Malaysia, Penang Malaysia 11700, E-mail: Muawya.cod08@student.usm.my }

\author{Pedro Peris-Lopez}
\address{Information Security and Privacy Lab, Delft University of Technology,  Mekelweg 4, 2628 CD, Delft, The Netherlands, Email: P.PerisLopez@tudelft.nl}

\author{Mohammd Rafie}
\address{School of Computer Sciences, Main Campus, Universiti Sains Malaysia, Penang Malaysia 11700, E-mail: Rafie@cs.usm.my }

\author{Jan van der Lubbe}
\address{Information Security and Privacy Lab, Delft University of Technology, Mekelweg 4, 2628 CD, Delft, The Netherlands, Email: J.C.A.vanderLubbe@tudelft.nl}

\begin{abstract}

In this paper, we analyze the security of an RFID  authentication protocol proposed by Liu and Bailey \cite{R6}, called Privacy and Authentication Protocol  (PAP), and show its vulnerabilities and faulty assumptions. PAP is a privacy and authentication protocol designed for passive tags. The authors claim that the protocol, being resistant to commonly assumed attacks, requires little computation and provides privacy protection and authentication. Nevertheless, we propose two traceability attacks and an impersonation attack, in which the revealing of secret information (i.e., secret key and static identifier) shared between the tag and the reader is unnecessary. Moreover, we review all basic assumptions on which the design of the protocol resides, and show how many of them are incorrect and are contrary to the common assumptions in RFID systems.

\end{abstract}

\begin{keyword}
RFID, authentication, privacy, cryptanalysis, impersonation attack, tractability attack


\end{keyword}

\end{frontmatter}


\section{Introduction}
\label{int}

Numerous RFID protocols have been proposed with the goal of providing secure contact between readers and tags over the open radio channel. Nevertheless, tags have severe limitations in terms of circuitry (gate equivalents), storage, and power consumption; thus, designing an efficient and secure mutual authentication protocol is still a great challenge. There are many security risks linked to RFID technology, and of these, privacy and tracking are the most important. We will explain these in detail. Nevertheless, readers are urged to consult \cite{R13,R14,R15,MitrokotsaRT-2008-iwrt} to have a complete overview of the risks and threats linked to RFID technology.

\begin{description}

\item[Privacy:] Tag content, which may include sensitive information, is revealed when insecure tags are interrogated by readers. Tags and readers should thus be authenticated to overcome this problem. However, readers are frequently not authenticated, and tags usually answer in a completely transparent and indiscriminate way.

\item[Tracking:] A problem closely related to privacy is tracking or violations of location privacy. Even if access of tag content were only allowed to authorized readers, non-tracking might still not be guaranteed. The answer provided by tags is usually a constant value (i.e., a static identifier). Under this assumption, an attacker will be able to establish an association between tags and their owners. Additionally, we can relax our conditions and assume that tags only contain product codes rather than unique identifiers. Nevertheless, Weis et al. claimed that tracking is still be possible by using a constellation of tags \cite{R13}.

\end{description}

Privacy and authentication protocol (PAP), like many proposed protocols \cite{R1,R3,R4,R7}, can be categorized under the class of a simple mutual authentication protocol \cite{R2}, in which tags can generate a random number and have computation recourses to compute a hash function. However, unlike other protocols in the area of simple mutual authentication protocol, PAP is designed in four sub-protocols based on the specific location of the tag.

For the rest of this paper, we first present a full review of PAP and illustrate the functionality of the four proposed sub-protocols. Then, we introduce and discuss the vulnerability analysis of the protocol and its assumptions categorized under conceptual and operational view, respectively. Thereafter, we propose two traceability attacks, followed by an impersonation attack. Finally, we offer some conclusions.

\section{Review of PAP}
\label{REV}

PAP claims to have a protocol, which is compliant with the Electronic Product Code Class-1 Generation-2 (EPC-C1G2) standard \cite{R11} or ISO/IEC 18006-C \cite{ISO18000-6} equivalently, in which the entities involved consist of tags, readers, and a back-end database. The basic communication of the protocol is between tags and readers, and the authors do not discuss the matter of reader-database communication. Furthermore, compared with the original EPC-C1G2 standard, it is assumed that RFID tags support on-chip, a secure hash function, to provide stronger security guarantees.

Each tag in PAP has a secret key $k$ shared with the reader, a generic name (i.e., product type), a static $ID$ (i.e., unique identifier), and a privacy bit that can be switched between 0 and 1 (the 0/1 value denotes that the tag is in a secure/insecure location). Readers are classified into three categories, namely, inventory, checkout, and return readers. The implementation scenario considers four locations for the tag to be in: 1) inside the store; 2) at the checkout counter; 3) outside the store; and 4) at the return counter. Each location has its own sub-protocol in PAP in terms of maintaining privacy and authentication for the tags all the time and at any location.

The basic difference among the four PAP sub-protocols is that the privacy bit value is equal to zero, as in the cases of in-store and checkout sub-protocols. Here, the tag is assumed to be inside the store and sends its $ID$ during communication with the reader. Otherwise, the generic name, which is a numeric representation of the product type, communicates with the reader when the privacy bit value is set to 1 as in the cases of the out-store and return sub-protocols.

The in-store and out-store sub-protocols both share the same exchanged messages. Basically, tags send back one message after being queried by a reader. However, the exception for the message is that the tag sends a static identifier $ID$ in the former and the tag sends a generic name in the latter. In both cases, a random number is sent in the tag response, but this does not have any security purpose. The in-store sub-protocol does not offer authentication or privacy protection because its main goal is efficiency. It assumes that the store provides other mechanisms to prevent malicious readers from entering. In the out-store sub-protocol, privacy protection is achieved by sending a generic name number, not the ID, where an unauthorized reader can read a tag but will not know what specific item is being read at that time.

The checkout sub-protocol and the return sub-protocol are sequels to the in-store and out-store sub-protocols, respectively; they also differ in terms of dealing with the $ID$ of the former and in the generic name of the latter. These protocols share the same exchanged messages to establish mutual authentication between the tag and the reader. The first two messages (i.e., those that are used in the in-store and out-store sub-protocols) are identical. Next, the reader retrieves the secret key $k$ from the database using the $ID$ or the generic name received from the tag. The reader computes a one-way hash function on $k$ and the random number $n_t$ received from the tag ($H_1 = hash(n_t, k)$).  It then computes a new nonce $n_r$ and sends both values ($H_1, n_r$) to the tag. The tag checks the hash value received. If so, it authenticates the reader and switches the privacy bit from 0 to 1. Finally, the tag computes $H_2 =hash(n_r,k)$ and forwards this value to the reader. Upon checking its correctness, the reader authenticates the tag. Figure (1) illustrates these four sub-protocols.

\section{Vulnerability Analysis of PAP}

The PAP threat model presents several assumptions that are considered essential for the protocol to ensure the privacy and authentication guarantees that it offers when applied in real-life scenarios. The authors partially justified these assumptions based on previous works, but they failed to mention these references. Furthermore, many of these assumptions are not compliant with the standard (EPC-C1G2 or ISO/IEC 18006-C) that the proposed protocol claims to be compatible with. For example, Chien \cite{R2} categorized RFID mutual authentication protocols into four classes, namely, full-fledged, simple, lightweight, and ultra lightweight. EPC-C1G2-friendly protocols support only the last two categories when PAP uses a hash function and a random number generator that are supported by the first two categories only.

The criticisms of these assumptions are summarized in two views: 1) conceptual view (to analyze the design incompatibilities) and 2) operation view (to examine inaccuracies within the implementation assumptions). We explain these views below.
\begin{description}
	\item [Conceptual view:] PAP aims to provide authentication and privacy for the tag when it resides inside or outside the store using the privacy bit. More precisely, the authors ensure these by not sending the tag's  $ID$ outside the store. When the realization of the privacy bit is technically straightforward, its activeness against rouge readers is questionable \cite{rug_red_1, rug_red_2}. PAP categorizes readers into three types, namely, inventory, checkout, and return readers. An inventory reader is capable of querying a tag, but is not connected to a database. Nevertheless, this assumption is mistaken because these readers will not be able to authenticate (legitimate or dishonest) tags and, therefore, counterfeit tags are never detected. This is claimed to be the only difference between the inventory reader and the other two discriminated readers. In fact, the other two readers have the same capabilities, but have different label names. On the other hand, PAP is disseminated into four sub-protocols. This dissemination is unnecessary when there is one protocol with different states. In-store and out-store sub-protocols are no more than the first part of the protocol. The return sub-protocol is a special case of the original checkout sub-protocol when the privacy bit value is equal to one and, accordingly, tags send a numerical generic name instead of the tag's $ID$. Finally, PAP aims to lessen the cost of the tags by sending the random value $n_t$ once at the beginning of every sub-protocol so that the tag would not need further programming for each sub-protocol. Although this might be true when using the four sub-protocols of PAP, it does not, however, reduce the initial cost of the tag. More precisely, it also does not justify the relatively high cost of implementing on-chip of low-cost tags a secure hash function \cite{FeldhoferR-2006-rfidsec,BogdanovLPPRS-2008-ches}.	

\item [Operation view:] Given that PAP was designed for RFID tags, capable of computing a secure hash function, PAP cannot be implemented on EPC-C1G2 tags, as claimed by the authors. Specifically, EPC-C1G2 low-cost tags do not support hash functions; it only supports simple operations, a 16-bit cyclic redundancy check, and a 16-bit pseudorandom number generator. Furthermore, the authors state that the short-range transmission between a reader and a tag provides a secure channel because an adversary cannot get in-between. This assumption is contrary to most studies on the security of channels, where it is commonly assumed that an adversary is capable of eavesdropping on the forward (reader-to-tag) and backward (tag-to-reader) channels \cite{R16,R17}. The authors misconstrue this assumption based on studies implemented on different types of tags, and mostly on different application areas or models than the ones proposed in the paper (e.g., place-on readers where the distance between the tagged object, such as contactless ID cards, and the reader allows no such space for malicious readers to be placed in the middle). Nevertheless, in the applications referred to in PAP, the nature of the objects holding the tags and real-life implementation scenarios cannot support this assumption. As such, the protocol presumes that the store ensures extra security measures to prevent any reader from getting inside the store, and that is unacceptable for public-related applications (e.g., Wal-Mart). It is also unrealistic to exclude eavesdropping attacks for the same reason, because an adversary could be outside the store and still be able to eavesdrop on a tag inside the store when the privacy bit value is zero. To the best of our knowledge, there are no proper ways to detect or prevent malicious readers from getting inside the store (remember, we are using non-secure and public radio channel). Furthermore, PAP claims that real-time message compromising (i.e., interception/alteration) is difficult and is therefore discarded. However, numerous studies \cite{R5,R8,R9,R10} considered it an effective threat in many applications, giving it an important role in protocol design.
\end{description}

\subsection{Tractability Attacks}

One of the main concerns linked to privacy is the location-privacy or tractability protection, which indicates that unauthorized readers are unable to identify the location of the tag (or the tag's holder) by tracing the tag's transmitted data. PAP uses strong cryptographic primitives on the tag, such as a pseudorandom number generator and a hash function, that are considered relatively costly (especially the latter) in terms of gate equivalents (circuit area), and whose support increases significantly in the final cost of the chip. The use of these primitives should offer a highly non-traceable protocol. Nevertheless, PAP does not fully utilize the computing capabilities that are supported on-chip of the tag and prefers efficiency over security in some of its sub-protocols (i.e., in-store and out-store sub-protocols). The check-out and return sub-protocols are designed weakly and are vulnerable to traceability attacks despite using random numbers and computing authentication tokens by running a hash function. The traceability deficiencies are described in the sections below.

\begin{enumerate}
	\item In the in-store and out-store protocols, the tag always sends fixed values corresponding to the tag's $ID$ and tag names, respectively. In the former, the tag's holder can be tracked because the $ID$ is unique and constant. In the latter, tag name is constant but not a unique value. Nevertheless, the tag's holder can be tracked using constellations of tag names that are unequivocally linked to a specific user. The use of the random numbers, $n_t$, avoids further programming in the different sub-protocols. However, in these sub-protocols, the $n_t$ value is not used for any security purpose. Furthermore, the security of these sub-protocols depends solely on the cover-coding protection mechanism, which is considered completely ineffective and insecure \cite{PerisHTR-book} compared with the use of the hash function -- specially when the hash function is used properly. More precisely, the use of a cover-coding mechanism does not protect the values transmitted by the tag (i.e., $ID$ or tag name). In fact, the tag is insecure and is traceable if an attacker can eavesdrop on the forward or backward channels, which is the most common assumption in RFID systems.

\item In the check-out and return sub-protocols, the adversary can track the tag by setting one of the random numbers used in the protocol (i.e. $n_t$ or $n_r$) to a constant value ``$c$''. There are two alternatives; we can track either the reader's answer or the tag's answer. In a forward-channel tractability attack, the adversary can intercept the tag's reply to the reader's query. Then, he/she replaces $n_t$ with constant value $c$, and finally forwards the message ($ID$/name, $c$) to the reader. The rest of the protocol would conclude normally. We emphasize here that the hashed value $H_1$ is the same every time the attack is executed because the adversary fixes $n_t$ to $c$ ($H_1=hash(c,k)$). If the adversary runs the attack twice, as illustrated in Figure (2), he/she is able to track the tag by checking the equality between the $H_1$ values. In the backward-channel tractability attack, the reader receives the tag's reply on its query, hashes $n_t$ ($H_1=hash(n_t, k)$), and sends the hashed value accompanied by a new random number $n_r$ to the tag ($H_1, n_r$). The adversary intercepts the message, replaces $n_r$ with a $c$ constant and forwards the message ($H_1$, $c$) to the tag. The rest of the protocol would conclude normally. As the adversary sets $n_r$ to $c$, the hashed value $H_2$ is the same every time the attack is executed, and the adversary is able to track the tag by checking the equality between $H_2$ values. Figure 3 shows the two executions of the described attack.
\end{enumerate}

\subsection{Impersonation Attacks}

PAP claims that an adversary cannot impersonate a tag because the value of the secret key $k$  is never sent over the insecure radio channel and that the adversary cannot compute the authentication tokens ($H_2=hash(n_r,k)$). However, an adversary can impersonate a tag using the answers provided by a second legitimate reader. More precisely, the adversary exploits the symmetry of the messages computed by the reader and the tag in the PAP. In other words, the values of $H_1=hash(n_t,k)$ and $H_2=hash(n_r,k)$  have the same structure. We describe the attack in detail below.

An impersonation attack can be conducted between an adversary and two legitimate readers. The second reader generates the messages of the supplanted tag. We assume that before launching the attack, the adversary eavesdrops on the $ID$/name of its target tag. First, reader$_1$ sends a request query to the adversary. The adversary replies with $\{ID$/name, $n_t\}$, where $n_t$ represents an arbitrary random value. Then, reader$_1$ computes its authentication token $H_1$, generates a random value $n_r$, and sends both values to the adversary. The adversary simulates that he/she received a request from reader$_2$  and sends the  $\{ID$/name, $n_r\}$. The adversary uses the random number $n_r$ received from reader$_1$. The reader$_2$ computes its authentication token ($H_1^*=hash(n_r,k)$) and sends it to the adversary. Finally, the adversary forwards $H_1^{*}$  value to reader$_1$. The reader$_1$ checks the token received and authenticates the adversary. So, impersonation is viable by forwarding messages of a second legitimate reader and without disclosing the secret key of the tag. The attack is illustrated in Figure 4.

\section{Conclusions}

We introduced a vulnerability analysis of PAP for RFID tags. As shown in the paper, the assumption that PAP is an EPC-C1G2-friendly protocol is not well founded due to the excessive resources needed to support a hash function on board. Therefore, the proposed protocol is not suitable for low-cost RFID tags.

The paper first scrutinized the basis on which the protocol has been founded and then showed how two of its main security objectives can be compromised. More precisely, the assumptions determining how PAP provides a secure (privacy and authentication) RFID protocol have been discussed and determined to be incorrectly assumed during the design of the protocol. Then, we analyzed the protocol for traceability attacks, after which we illustrated respective attacks on the forward and backward channels. Finally, we conducted an impersonation attack by forwarding the messages computed by a second legitimate reader.
In summary, protocol designers should check carefully a protocol's compatibility with standards (e.g., EPC-C1G2 or ISO/IEC 18006-C). Moreover, the design of a secure and efficient RFID authentication protocol remains to be a challenge and not a completely resolved issue.

\begin{figure}[htbp]
\centering
	\includegraphics[scale=0.4]{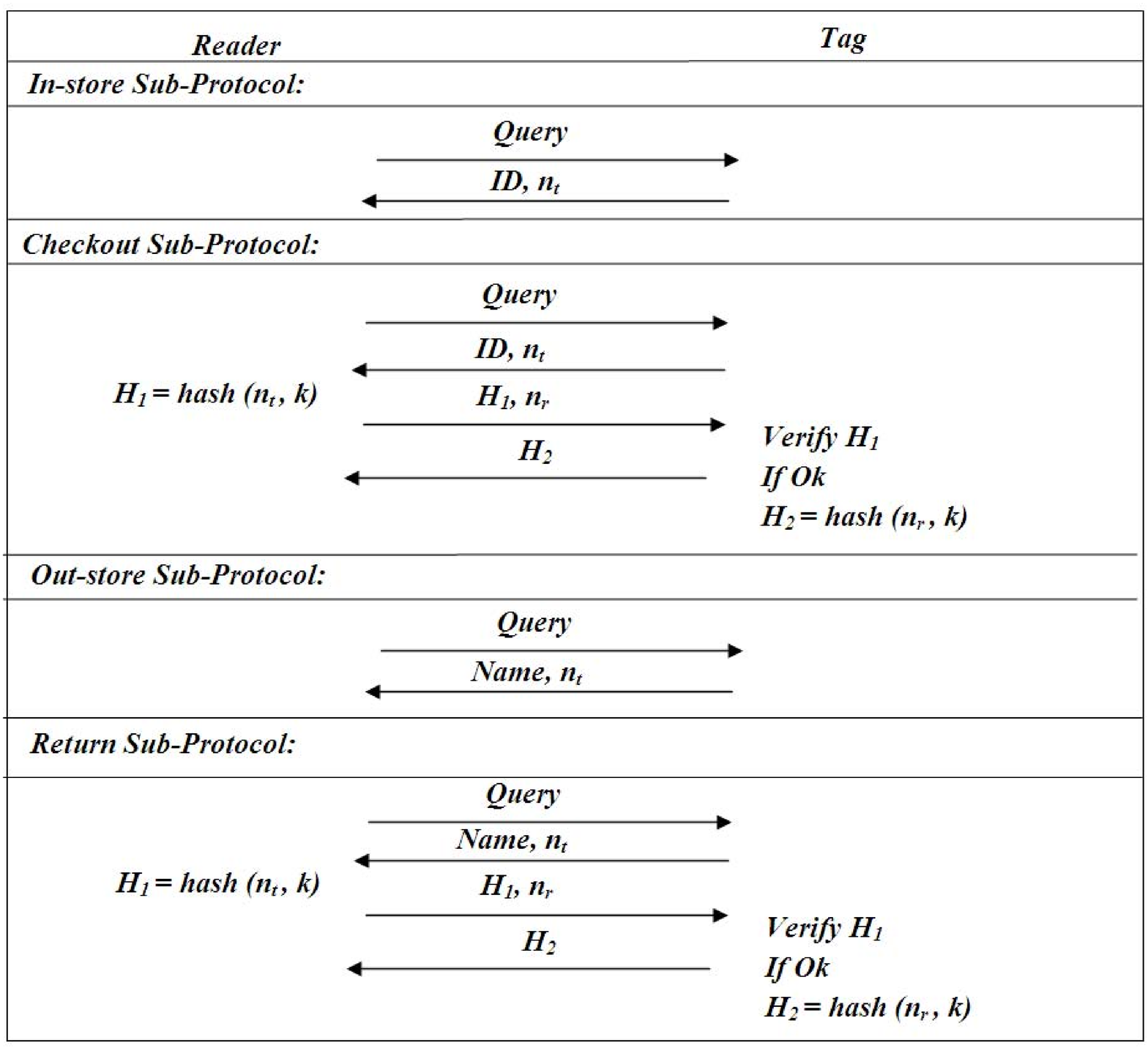}
	\caption{PAP sub-protocols}
	\label{fig:PAPProtocols}
\end{figure}

\begin{figure}[htbp]
\centering
	\includegraphics[scale=0.3]{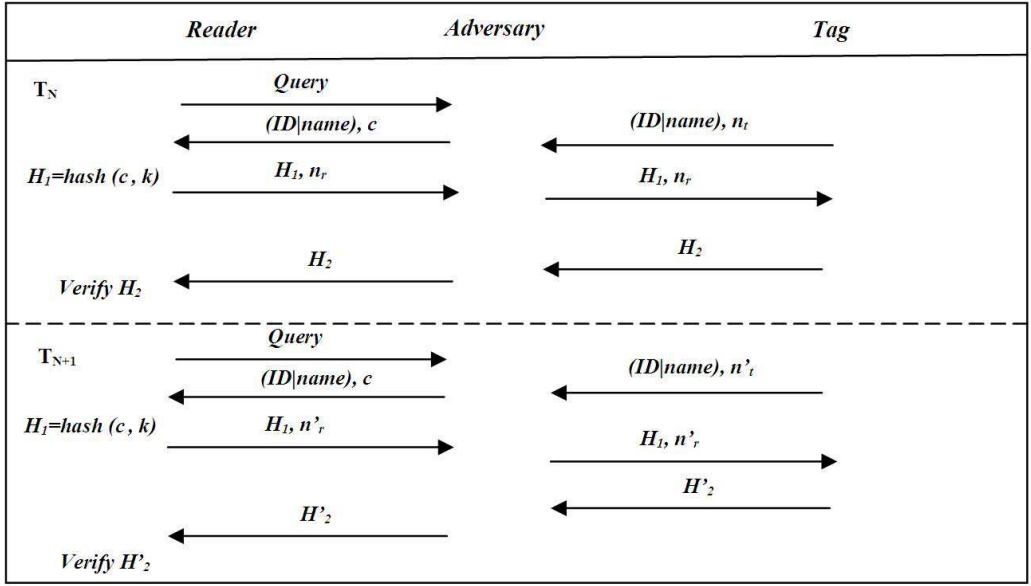}
	\caption{Forward-channel tractability attack}
	\label{fig:PAPProtocols}
\end{figure}

\begin{figure}[htbp]
\centering
	\includegraphics[scale=0.3]{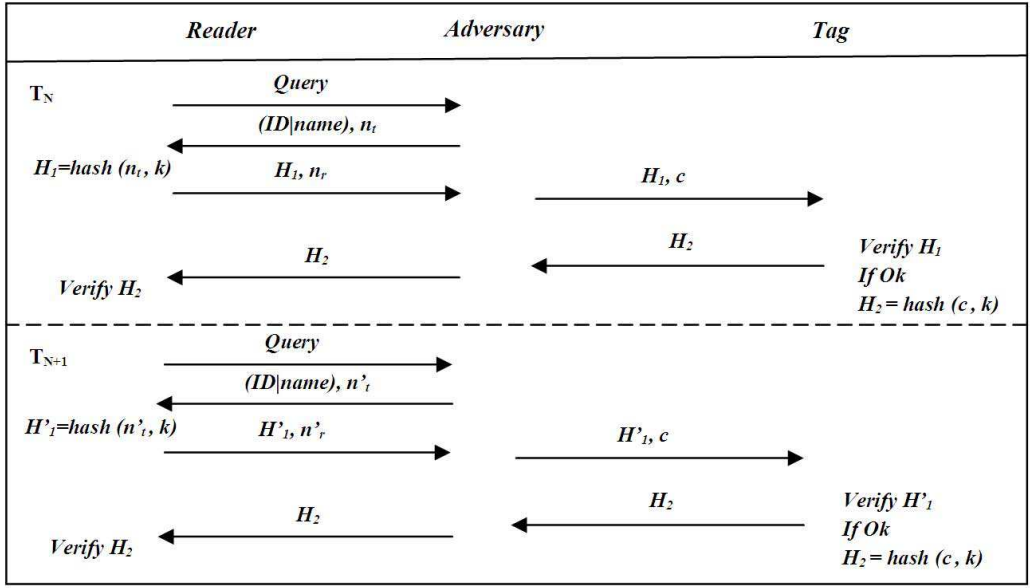}
	\caption{Backward-channel tractability attack}
	\label{fig:PAPProtocols}
\end{figure}

\begin{figure}[htbp]
\centering
	\includegraphics[scale=0.3]{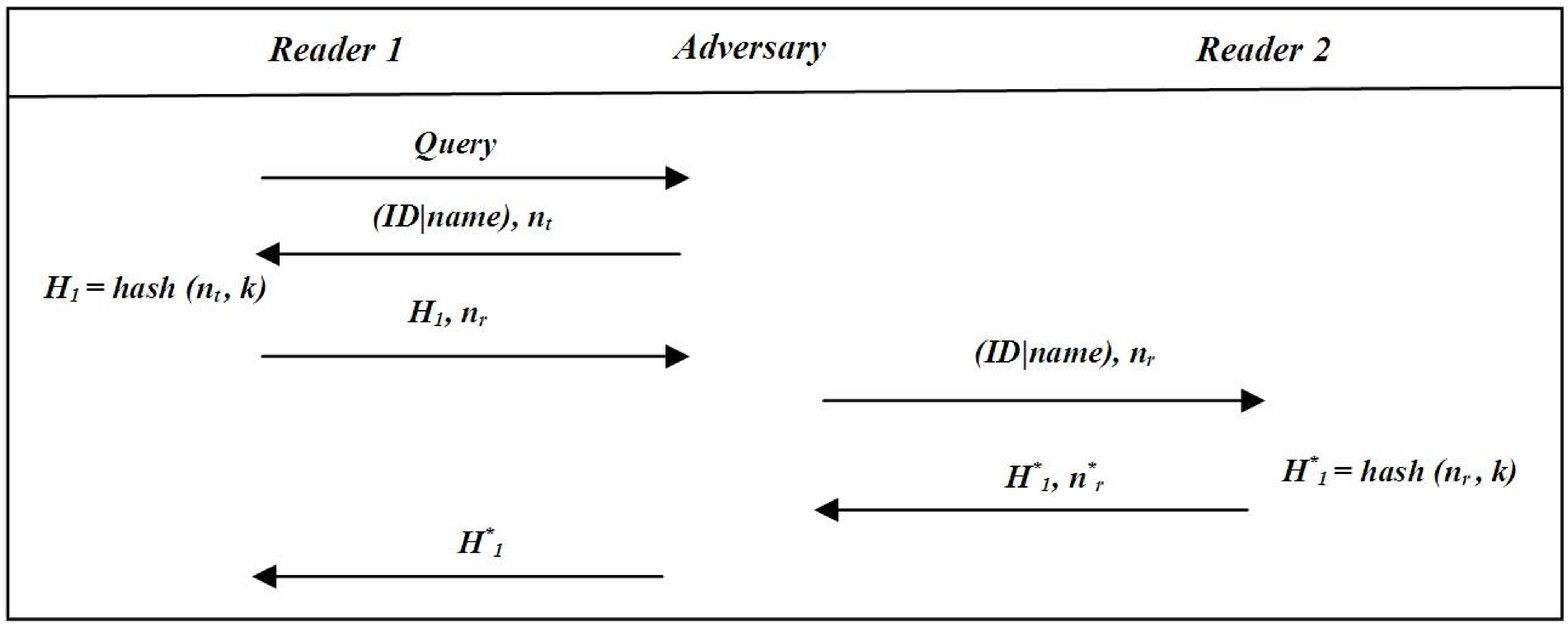}
	\caption{Impersonation  attack }
	\label{fig:PAPProtocols}
\end{figure}

\end{document}